\documentclass[twocolumn]{aastex62}

\submitjournal{ApJ}

\shorttitle{Cosmic-Ray Escape from the SMC}
\shortauthors{Lopez et al.}

\newcommand{\ltsima}{$\; \buildrel < \over \sim \;$}
\newcommand{\simlt}{\lower.5ex\hbox{\ltsima}}

\newcommand{\ls}{{_<\atop^{\sim}}}

\def\arcdeg{\hbox{$^\circ$}}
\def\arcmin{\hbox{$^\prime$}}

\begin{document}

\title{Evidence for Cosmic-Ray Escape in the Small Magellanic Cloud using {\it Fermi} Gamma-Rays}

\correspondingauthor{Laura A. Lopez}
\email{lopez.513@osu.edu}

\author{Laura A. Lopez}
\affil{Department of Astronomy, The Ohio State University, 140 W. 18th Ave., Columbus, Ohio 43210, USA}
\affil{Center for Cosmology and AstroParticle Physics, The Ohio State University, 191 W. Woodruff Ave., Columbus, OH 43210, USA}
\affil{Niels Bohr Institute, University of Copenhagen, Blegdamsvej 17, 2100 Copenhagen, Denmark}

\author{Katie Auchettl}
\affil{Center for Cosmology and AstroParticle Physics, The Ohio State University, 191 W. Woodruff Ave., Columbus, OH 43210, USA}
\affil{Department of Physics, The Ohio State University, 191 W. Woodruff Ave., Columbus, OH 43210, USA}

\author{Tim Linden}
\affil{Center for Cosmology and AstroParticle Physics, The Ohio State University, 191 W. Woodruff Ave., Columbus, OH 43210, USA}
\affil{Department of Physics, The Ohio State University, 191 W. Woodruff Ave., Columbus, OH 43210, USA}

\author{Alberto D. Bolatto}
\affil{Department of Astronomy, University of Maryland, College Park, MD 20742, USA}

\author{Todd A. Thompson}
\affil{Department of Astronomy, The Ohio State University, 140 W. 18th Ave., Columbus, Ohio 43210, USA}
\affil{Center for Cosmology and AstroParticle Physics, The Ohio State University, 191 W. Woodruff Ave., Columbus, OH 43210, USA}

\author{Enrico Ramirez-Ruiz}
\affil{Niels Bohr Institute, University of Copenhagen, Blegdamsvej 17, 2100 Copenhagen, Denmark}
\affil{Department of Astronomy \& Astrophysics, University of California, Santa Cruz, CA 95064, USA}

\begin{abstract}

Galaxy formation simulations demonstrate that cosmic-ray (CR) feedback may be important in the launching of galactic-scale winds. CR protons dominate the bulk of the CR population, yet most observational constraints of CR feedback come from synchrotron emission of CR electrons. In this paper, we present an analysis of 105~months of {\it Fermi} Gamma-ray Space Telescope observations of the Small Magellanic Cloud (SMC), with the aim of exploring CR feedback and transport in an external galaxy. We produce maps of the $2-300$ GeV emission and detect statistically significant, extended emission along the ``Bar" and the ``Wing", where active star formation is occurring. Gamma-ray emission is not detected above $\sim$13~GeV, and we set stringent upper-limits on the flux above this energy. We find the best fit to the gamma-ray spectrum is a single-component model with a power-law of index $\Gamma=-2.11\pm0.06\pm0.06$ and an exponential cutoff energy of $E_{\rm c} =13.1\pm5.1\pm1.6$~GeV. We assess the relative contribution of pulsars and CRs to the emission, and we find that pulsars may produce up to 14$^{+4}_{-2}$\% of the flux above 100~MeV. Thus, we attribute most of the gamma-ray emission (based on its spectrum and morphology) to CR interactions with the ISM. We show that the gamma-ray emissivity of the SMC is five times smaller than that of the Milky Way and that the SMC is far below the ``calorimetric limit", where all CR protons experience pion losses. We interpret these findings as evidence that CRs are escaping the SMC via advection and diffusion.

\end{abstract}

\keywords{cosmic rays --- Magellanic Clouds --- gamma rays: ISM}

\section{Introduction} 

Cosmic rays (CRs) have a profound influence on the interstellar medium (ISM) and in galaxies (see reviews by \citealt{strong07}, \citealt{zw13}, and \citealt{grenier15}). In the Milky Way (MW), CRs play a fundamental role in the ISM, contributing equal pressure as the magnetic field, turbulence, radiation, and thermal components (e.g., \citealt{boulares90}). CRs are the primary ionization mechanism of molecular gas (which is shielded from UV photons; \citealt{dalgarno06}), and CRs are responsible for the production of light elements (Li, Be, and B) via spallation of O and N atoms  \citep{fields99,fields00,ramaty00}. On galactic scales, CRs may be important in launching winds (e.g., \citealt{ipavich75,breit91,zirak96,ptuskin97,everett08,socrates08,samui10,dorfi12,uhlig12,gir16}). 

Substantial attention has been devoted to incorporate CR feedback into galaxy formation simulations (e.g., \citealt{jubelgas08,booth13,salem14,salem14b,pakmor16,rus16,simpson16,pfrommer17a,wiener17,jacob18}). These works vary in how they model CR transport (e.g., isotropic diffusion, anisotropic diffusion, advective streaming), and the results show that galactic wind properties (e.g., mass loading, velocity) differ depending on their assumptions (though see \citealt{pfrommer17b}).

In order to model CRs properly, it is vital to observe how CRs are transported within a variety of galaxies and conditions. Historically, the primary means to probe CRs in other galaxies is through study of the radio emission from CR electrons. For example, the tight correlation between galaxies' far-infrared (FIR) luminosity (a tracer of massive star formation: \citealt{kenn12}) and their synchrotron radiation (associated with GeV CR electrons) in the radio (e.g., \citealt{helou85,condon92,yun01}) supports an intrinsic connection between star formation and CRs. 

Advances in GeV and TeV astronomy, with facilities like the {\it Fermi} Gamma-ray Space Telescope \citep{atwood09} and the High Energy Stereoscopic System (H.E.S.S.: \citealt{hinton04}), enable spatially-resolved studies of gamma-rays from CR protons, which comprise the bulk of the CR population. In particular, CR protons interacting with dense gas produce pions, which decay into gamma-rays that dominate the spectrum $\sim$0.1--300~GeV in star-forming galaxies (e.g., in the MW: \citealt{strong10}). {\it Fermi} studies of the integrated GeV emission from star-forming galaxies show a FIR/gamma-ray correlation similar to the FIR/radio correlation \citep{ackermann12}. 

As the nearest star-forming galaxies to the MW, the Magellanic Clouds are resolved and detected at GeV energies with {\it Fermi} \citep{abdo10,abdo10b}. Using 17~months of {\it Fermi} Large Area Telescope (LAT) observations, \cite{abdo10b} reported the initial detection of the Small Magellanic Cloud (SMC) at $\sim$11-$\sigma$ significance, and they modeled the emission as an extended source with a $\sim$3$^{\circ}$ diameter. However, no substructure was readily apparent in the data, and they noted the emission was not clearly correlated with the distribution of massive stars or neutral gas. They found that the observed flux of the SMC implies an average density of CR nuclei that is only $\sim$15\% of the value of the MW. Given that the CR injection rate of the SMC seems comparable to the MW, the authors concluded that the difference may be due to CR transport effects, such as the SMC having a smaller confinement volume. 

More recently, \cite{caputo16} analyzed six years of data from {\it Fermi}/LAT toward the SMC to search for gamma-ray signals from dark matter annihilation. They tested several different spatial templates, including a single two-dimensional Gaussian model and an emissivity model, which assumes the gamma-ray emission arises from cosmic rays interacting with interstellar gas. \cite{caputo16} found that both models did comparably well in  describing the extent of the SMC's gamma-ray emission. 

In this paper, we present a new analysis of 105~months (8.75~years) of {\it Fermi}/LAT data available from the SMC, a $\approx$5.5~times deeper integration than presented in \cite{abdo10b}. We exploit the improved effective area and resolution of the Pass~8 data to produce new gamma-ray maps and spectra of the SMC. In particular, we focus our imaging analysis on the data $\gtrsim$2 GeV to exploit the vastly improved spatial resolution of LAT at high energies (e.g., the 68\% containment radius at $\gtrsim$2 GeV is $\ls$0.1$^{\circ}$ compared to $\sim2^{\circ}$ at 200 MeV: \citealt{atwood09}). Using this approach, we resolve substructure in the SMC as well as the Galactic globular cluster NGC~362, which is $<1^\circ$ north of the SMC's star-forming Bar. The detection of NGC~362 adds it to a growing list of globular clusters that have been detected with {\it Fermi} (e.g., \citealt{abdo_gc,hooper16}), and these data are a useful tool to assess the millisecond pulsar population in globular clusters.

The paper is structured as follows. In Section~\ref{sec:data}, we describe the observations and analysis to produce the spectra and images of the SMC. In Section~\ref{sec:results}, we present the results for the SMC and for the Galactic globular cluster NGC~362. In Section~\ref{sec:discussion}, we discuss the implications for CR transport in the SMC, and Section~\ref{sec:conclusions} presents our conclusions. Throughout this paper, we assume a distance to the SMC of 61~kpc \citep{hilditch05}.

\section{Observations and Data Analysis} \label{sec:data}

Photon and spacecraft data from 105~months of observations with the {\it Fermi}/LAT (spanning from 4 August 2008 to 22 May 2017\footnote{Mission elapsed time (MET) range of 239557417 to 517134927}) were downloaded from the {\it Fermi} Science Support Center for the SMC (centered at right ascension $\alpha =$15.116$^{\circ}$ and declination $\delta =-72.966^{\circ}$) in a 30 degree region of interest (ROI). Pass~8 data were analyzed using {\it Fermi} Science Tools v10r0p5\footnote{The Science Tools package and support documents are distributed by the Fermi Science Support Center and can be accessed at http://fermi.gsfc.nasa.gov/ssc}. We used the ``P8R2\_SOURCE\_V6" instrument response function (IRF), and we selected events with a zenith angle $<$90$^{\circ}$ and cut those detected when the rocking angle was $>$52$^{\circ}$ to minimize contamination from the Earth limb. 

We use a maximum likelihood method to quantitatively explore the observed gamma-ray emission. Given a specific model for the distribution of gamma-ray sources on the sky and their spectra, the {\it Fermi} Science Tool command {\it gtlike} computes the best-fit parameters by maximizing the joint probability of obtaining the observed data from the input model. In this analysis, the likelihood $L$ is the probability that our spatial and spectral model represents the data, and the test statistic (TS) is defined as ${\rm TS} \equiv -2 {\rm log} (L_{\rm 0} / L_{\rm 1})$, where $L_{\rm 0}$ and $L_{\rm 1}$ are the likelihoods without and with the addition of a point source at a given position, respectively.

We performed a binned likelihood analysis using \hbox{{\it gtlike}} over the energy range of 200~MeV to 300~GeV. In our spatial and spectral analysis, we included all background sources from the LAT 4-year Point Source Catalog (3FGL: \citealt{3FGL}) within 20$^{\circ}$ of the SMC. Free parameters in the fit were the normalization of the Galactic diffuse emission, isotropic component, and background sources within 5$^{\circ}$ of the SMC. The normalizations of sources with angular distances of  $>5^{\circ}$ from the SMC were frozen to the values listed in the 3FGL. Instead of the spatial model of the SMC given in the 3FGL, we substituted a two-dimensional Gaussian (2DG) function centered at $\alpha=14.2^{\circ}$ and $\delta = -72.8^{\circ}$ with a width $\sigma=0.8^{\circ}$, as \cite{caputo16} reported an improved maximum likelihood with this model relative to that of the 3FGL. 

We also tried spatial models using multiwavelength images of the SMC as templates (e.g., H{\sc i}: \citealt{stan99}; H$\alpha$: \citealt{smith98}; H$_{2}$: \citealt{jameson16}; 70 $\mu$m: \citealt{gordon11}); all of these models were less successful than the 2DG reported by \cite{caputo16}. We note that the {\it Fermi}/LAT background model of the Galactic interstellar emission \citep{acero16}\footnote{https://fermi.gsfc.nasa.gov/ssc/data/access/lat/ \\ BackgroundModels.html} may include some contamination from the SMC, as a small positive residual is coincident with the western part of the SMC's Bar in {\it gll\_iem\_v06.fits}. If part of the SMC's emission is accounted for in the Galactic diffuse map, then the spatial modeling of the SMC may be affected. Given that it produces the best fit, we adopt the 2DG spatial model for the SMC in the subsequent analyses of this paper.

In addition, we added a point source to our background model at the position of $\alpha=5.9^{\circ}$ and $\delta = -68.3^{\circ}$, a source for which \cite{caputo16} derived a TS of 25--35, depending on the spatial model of the SMC. In this work, we find that this point source has a TS of 27.  Additionally, as discussed in Section~\ref{sec:ngc362}, a second point source was added at the position of $\alpha=15.65^{\circ}$ and $\delta = -70.94^{\circ}$, corresponding to the Milky Way globular cluster NGC~362 and yielded a TS of 32. 

\begin{figure*}[t]
\begin{center}
\includegraphics[width=\textwidth]{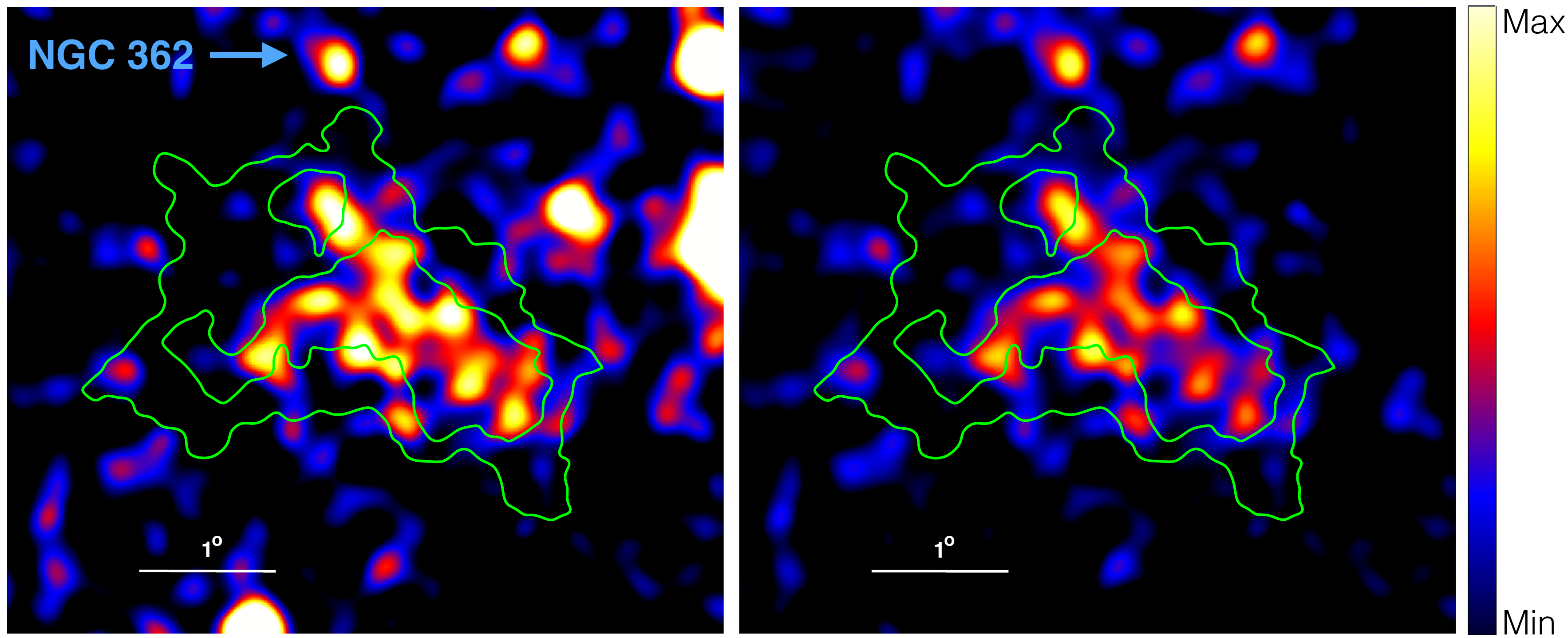}
\end{center}
\vspace{-3mm}
\caption{Total count map (left) and background-subtracted count map (right) of the SMC in the $2-300$ GeV band. The images were smoothed with a Gaussian function of width $\sigma = 10$~pixels = 6\arcmin. The green contour represents the distribution of H {\sc i} to guide the eye, the white scale bars denote 1$^{\circ}$. The color map is normalized to the maxima of the images. The position of the globular cluster NGC~362 is identified. North is up; East is left.}
\label{fig:fermi_image}
\end{figure*} 

To determine the detection significance and position of the SMC gamma-ray emission, we produced a TS map toward the SMC using the {\it Fermi} Science Tool command {\it gttsmap}, which computes the improvement of the likelihood fit when a point source is added to each finely-gridded spatial bin. We adopted the best-fit model output using {\it gtlike}, but removed the SMC and computed the TS value for 0.05$^{\circ}$ pixels across 5$^{\circ}$ centered on the galaxy. To maximize the spatial resolution of the data, we used only the $2-300$~GeV band to produce images. 

To produce the gamma-ray spectrum of the SMC, we use events converted in the {\it front} and {\it back} sections of the LAT with an energy range of $0.2-200$~GeV. We select this band to avoid the large uncertainties in the Galactic background model below 0.2 GeV. We model the flux in each of eight logarithmically-spaced energy bins and estimate the best-fit parameters using {\it gtlike}.  In addition to statistical uncertainties obtained from the likelihood analysis, systematic uncertainties associated with the Galactic diffuse emission were evaluated by altering the normalization of this background by $\pm6\%$ from the best-fit value at each energy bin (similar to \citealt{castro10} and \citealt{castro12})\footnote{https://fermi.gsfc.nasa.gov/ssc/data/analysis/LAT\_caveats.html}.

\begin{figure}
\begin{center}
\includegraphics[width=\columnwidth]{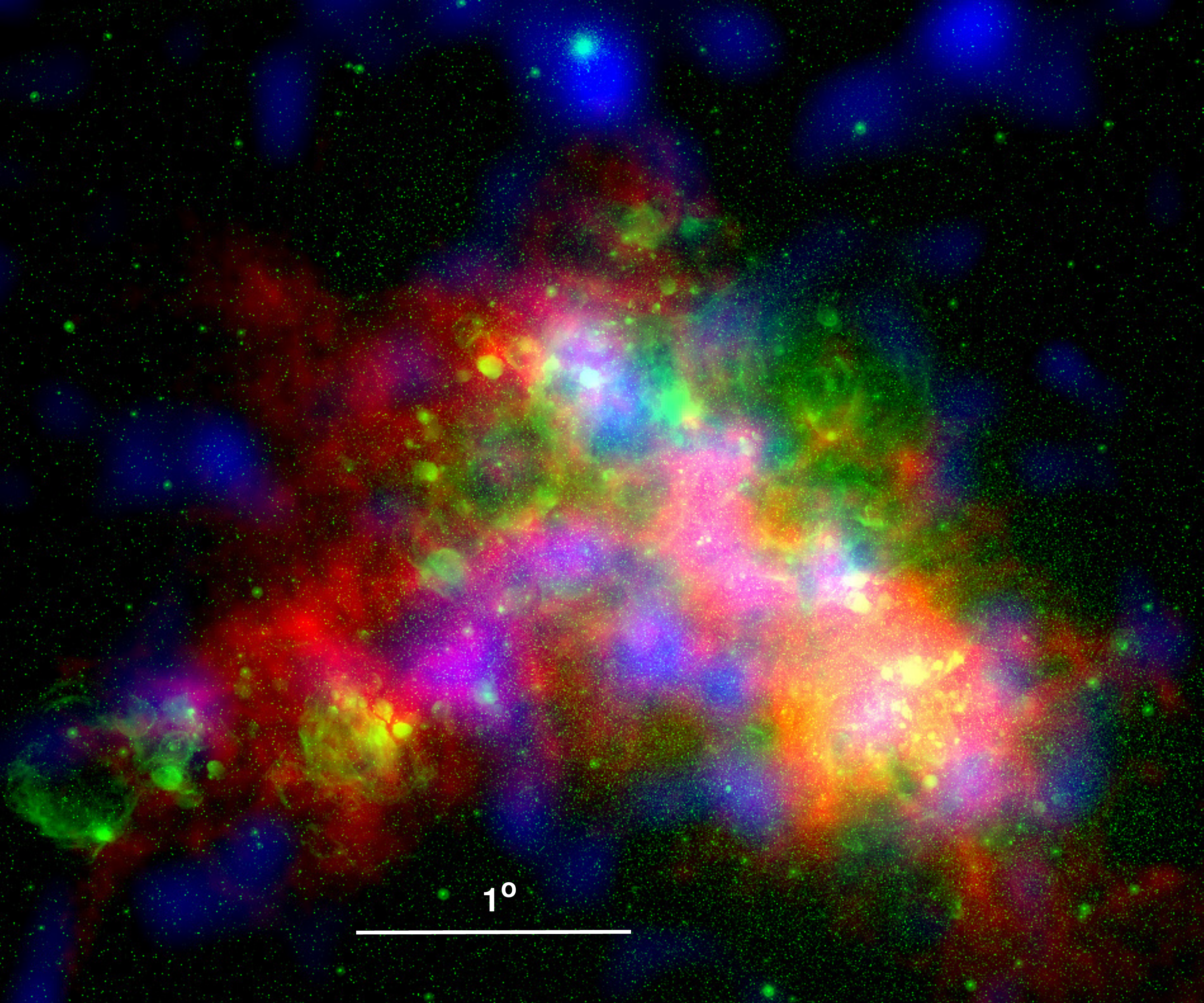}
\end{center}
\vspace{-3mm}
\caption{Three-color image of the SMC, with H{\sc i} in red, H-$\alpha$ in green, and the background-subtracted $2-300$~GeV count map in blue. The GeV gamma-rays show substructure along the ``Bar'' and ``Wing'' of the SMC and are largely coincident with the H{\sc i} and H-$\alpha$ emission. Additionally, the $2-300$ GeV gamma-rays are detected with statistical significance from the Galactic globular cluster NGC~362 (as seen at the top of the image). North is up; East is left.}
\label{fig:threecolor}
\end{figure} 

\section{Results} \label{sec:results}

\subsection{SMC} \label{sec:SMC}

Figure~\ref{fig:fermi_image} gives the $2-300$ GeV count map of the SMC before (left panel) and after (right panel) background subtraction\footnote{To produce the background-subtracted image, we generated an image at $2-300$~GeV using the {\it Fermi} Science Tool command {\it gtmodel} of all background sources using the best-fit parameters output by {\it gtlike}.}. The SMC is detected with 33.0-$\sigma$ significance in the $0.2-300$~GeV band. We compare the background-subtracted $2-300$~GeV count map to the H{\sc i} and H-$\alpha$ images of the SMC in Figure~\ref{fig:threecolor}. The gamma-ray emission has evident substructure: it is predominantly extended along the ``Bar'' of the SMC, where the bulk of the star formation is occurring \citep{kennicutt95,bolatto07}. Additionally, GeV gamma-rays are also detected in the direction toward the ``Wing'' of the SMC, to the southeast from the Bar. 

\begin{figure}
\begin{center}
\includegraphics[width=0.9\columnwidth]{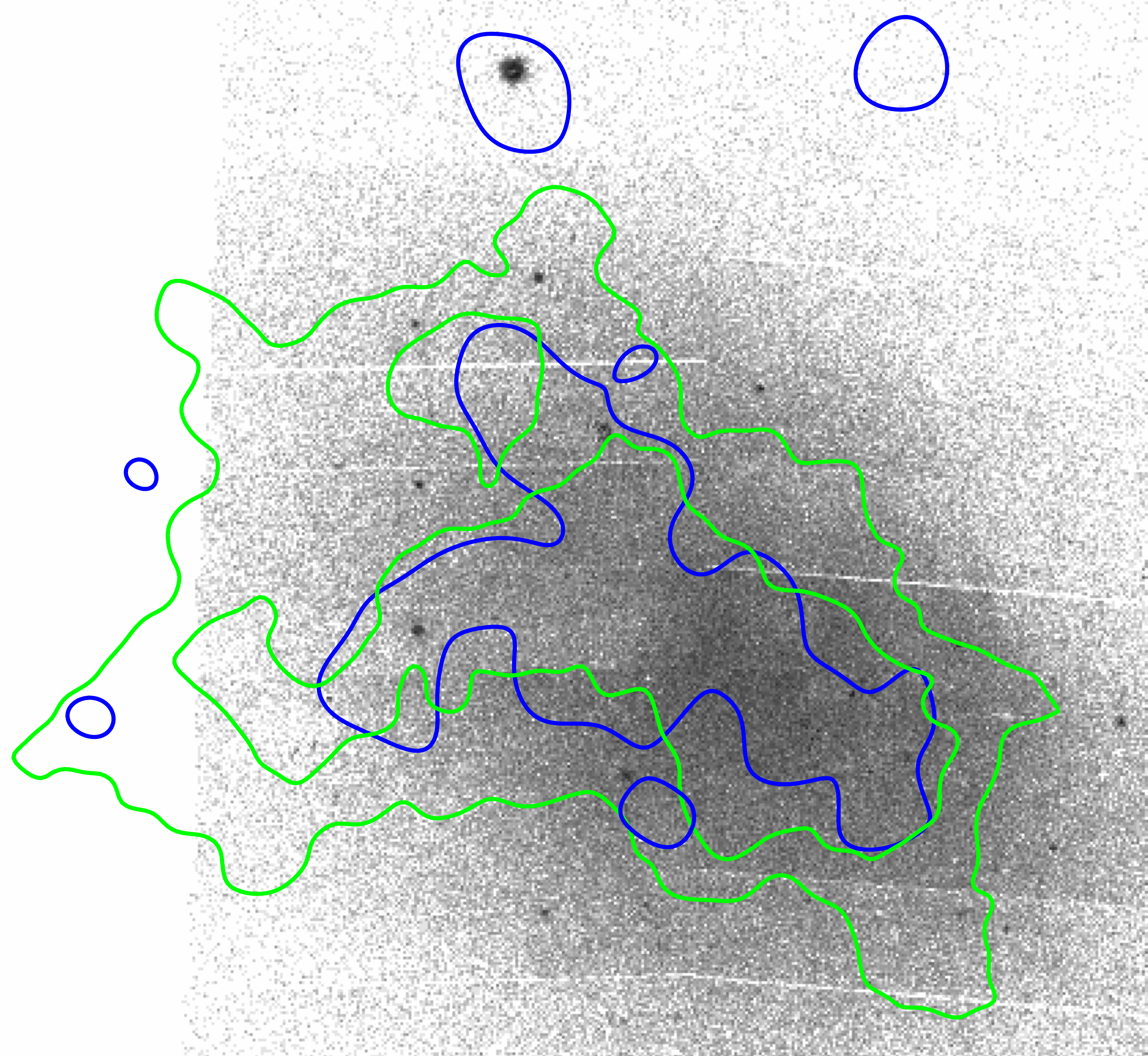}
\end{center}
\vspace{-3mm}
\caption{Stellar density map of the red giant and red clump stars (with ages $\gtrsim$1~Gyr) in the SMC \citep{zaritsky00,zaritsky02}, with green and blue contours representing the H{\sc i} and the 2--300~GeV gamma-rays, respectively. The old stellar population is distributed uniformly, whereas the H{\sc i} and gamma-rays are concentrated to the Bar and Wing of the SMC.}
\label{fig:oldstars}
\end{figure}

Figure~\ref{fig:oldstars} compares the gamma-ray distribution to that of the old stars in the SMC, specifically the stellar density map of red giant and red clump stars (with ages $\gtrsim$1~Gyr)\footnote{To generate the map of the old stars, we utilized the SMC stellar catalog from \cite{zaritsky02} and selected the red giant and red clump stars as those with $m_{\rm V} < 19.5$, $M_{\rm V} < 0.6$, and $B-V > 0.7$ \citep{zaritsky00}.}. The old stars have a fairly homogeneous distribution across the SMC, whereas the gamma-rays have evident substructure that follow the Bar and Wing morphology of the star-forming gas.

\begin{figure}
\begin{center}
\includegraphics[width=\columnwidth]{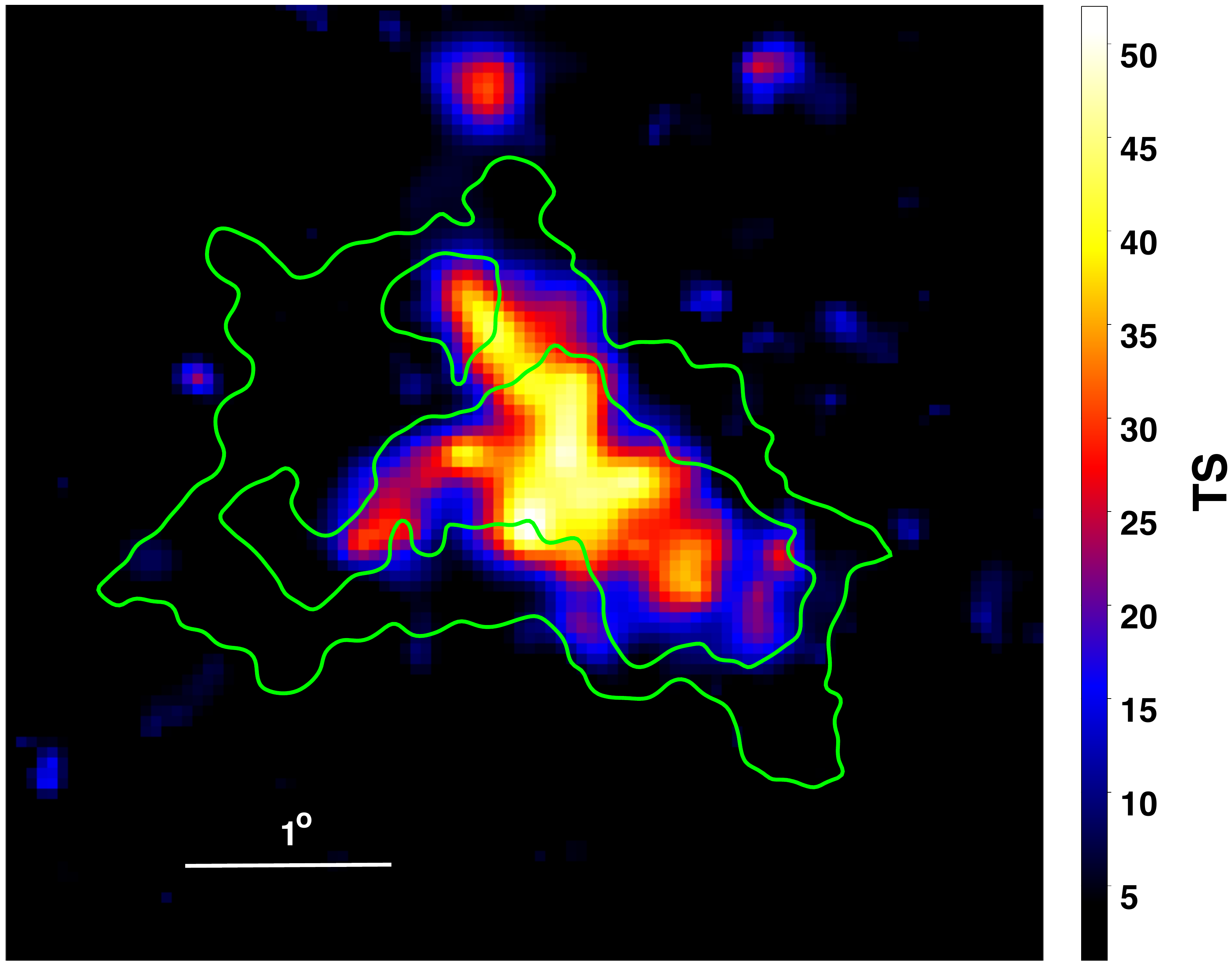}
\end{center}
\vspace{-3mm}
\caption{The TS map of the SMC in the $2-300$ GeV band. Values of TS$=$9 and TS$=$25 correspond to 3 and 5$\sigma$ detections, respectively. Most of the SMC Bar is detected with statistical significance, up to $\sim$7$\sigma$ in each 0.05$^{\circ}$ pixel. Gamma-ray extension along the SMC Wing is also detected with $\sim$5$\sigma$ significance. The $\sim$5$\sigma$ detection north of the SMC Bar is coincident with the Galactic globular cluster NGC~362. The green contours represent the distribution of H {\sc i}, and the white scale bar denotes 1$^{\circ}$. North is up, and East is left.}
\label{fig:tsmap}
\end{figure}

\begin{figure*}
\begin{center}
\includegraphics[width=\textwidth]{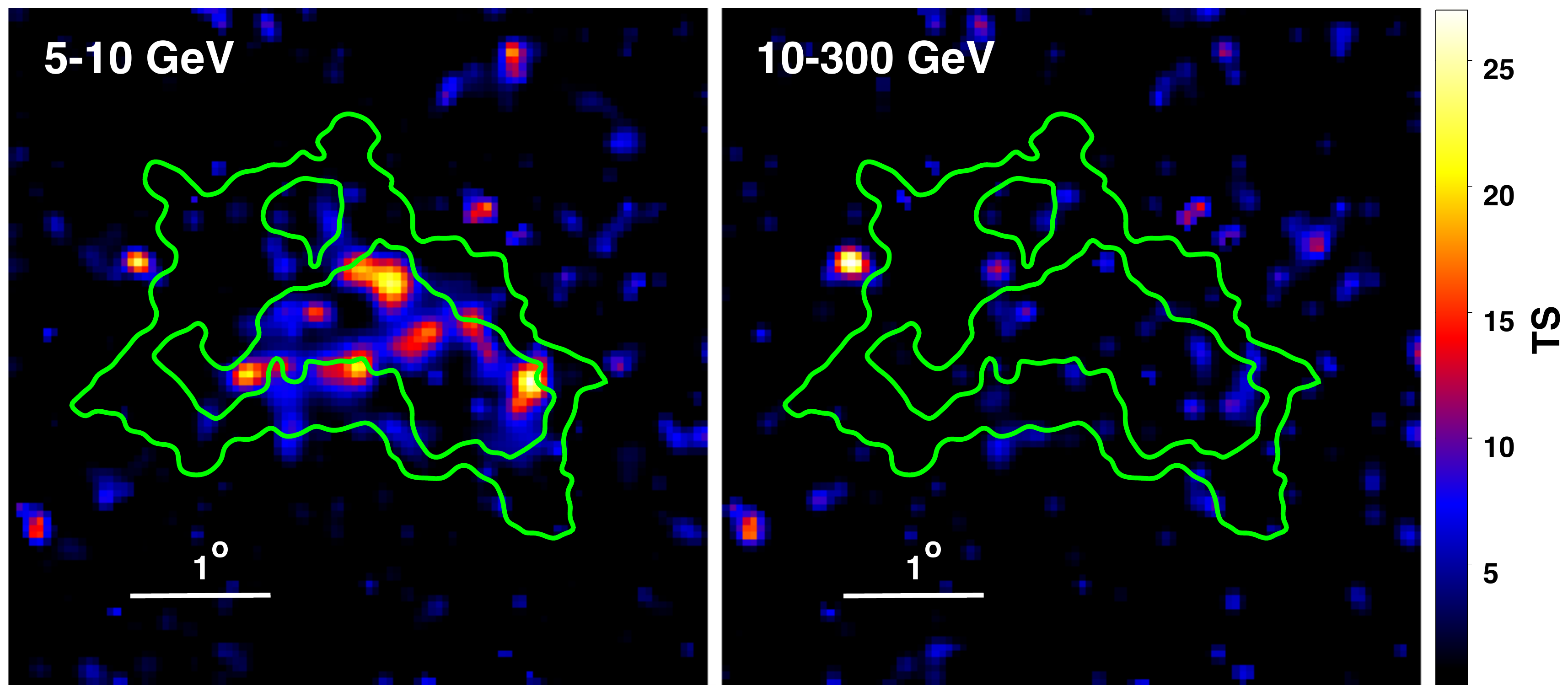}
\end{center}
\vspace{-3mm}
\caption{TS maps of the SMC in the $5-10$~GeV band (left) and the $10-300$ GeV band (right). Values of TS$=$9 and TS$=$25 correspond to 3 and 5$\sigma$ detections, respectively. Several locations in the SMC Bar and Wing are detected with $\sim$5$\sigma$ significance in the $5-10$ GeV band, whereas the regions have $\lesssim$3$\sigma$ significance emission in the $10-300$ GeV band. The green contours represent the distribution of H {\sc i}, and the white scale bar denotes 1$^{\circ}$. North is up, and East is left.}
\label{fig:tsmaps_higher}
\end{figure*}

\begin{figure}
\begin{center}
\includegraphics[width=\columnwidth]{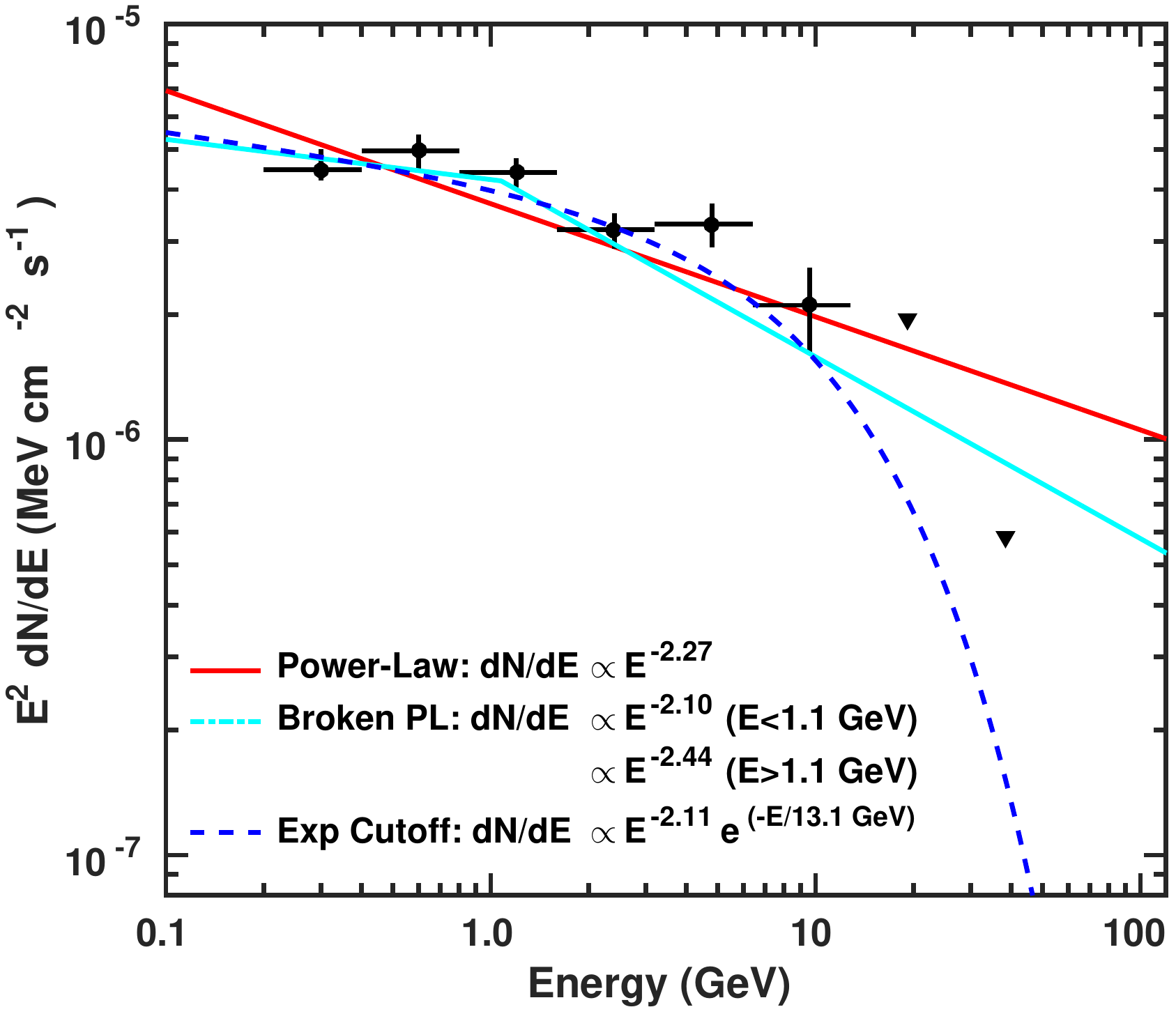}
\end{center}
\vspace{-3mm}
\caption{{\it Fermi}/LAT gamma-ray spectrum of the SMC. Error bars represent the statistical and systematic uncertainties, and the upper limits are shown as upside-down triangles. Overplotted are the best fits, using a simple power-law model or a power-law with an exponential cutoff. The upper limits in the $>$13~GeV bins are sufficiently stringent that the exponential cutoff model provides a statistically better fit.}
\label{fig:spectra}
\end{figure}

Figure~\ref{fig:tsmap} shows the TS map derived toward the SMC in the $2-300$ GeV band. We find that the majority of the SMC Bar and extension toward the Wing represent statistically significant detections (with TS$=$9 and TS$=$25 signifying 3- and 5-$\sigma$ detections, respectively) in the $2-300$~GeV band. When limited to $5-10$ GeV, the emission is concentrated in discrete locations of the SMC Bar and Wing (see Figure~\ref{fig:tsmaps_higher}), with multiple regions detected above 3-$\sigma$ confidence. We find no statistically significant signal (with TS$>$9) in the $10-300$~GeV band, consistent with the spectrum shown in Figure~\ref{fig:spectra}.

Figure~\ref{fig:spectra} gives the integrated gamma-ray spectrum of the SMC, with the statistical and systematic errors plotted for each data point. Photons $\gtrsim$13~GeV were not detected, so 2-$\sigma$ upper limits were determined for the two highest-energy bins ($12.8-25.6$ GeV and $25.6-51.2$ GeV). We plot the best-fit power-law (PL), broken power-law (BPL), and exponentially cutoff power-law (ECPL) models (see their functional forms in the legend of Figure~\ref{fig:spectra}), and the best-fit parameters and fluxes for each model are listed in Table~\ref{table:spectra}. Both the BPL and ECPL models are better at describing the data than the single PL model, with an improvement of $\Delta$TS$\approx$8.4 and $\Delta$TS$\approx$11.4, respectively. The ECPL model is statistically the best fit to the data and yielded a spectral index of $\Gamma = -2.11\pm0.6\pm0.6$ and a cutoff energy of $E_{\rm c} = 13.1\pm5.1\pm1.6$~GeV. We note that in their analysis of 17~months of {\it Fermi} data toward the SMC, \cite{abdo10b} found that an ECPL model (with a $\Gamma = -1.76^{+0.22-0.00}_{-0.14-0.01}$ and cutoff energy of $E_{\rm c} = 3.8^{+3.6+1.8}_{-1.3-0.8}$~GeV) fit the data better than a simple PL model with 2.4-$\sigma$ significance. \cite{caputo16} reached a similar conclusion using 6~years of {\it Fermi} data, though they reported a larger best-fit cutoff energy of $E_{\rm c} = 8\pm4$~GeV. Thus, the deeper {\it Fermi} analyzed here verifies with statistical significance of 3.4-$\sigma$ that the SMC spectrum appears to steepen or cutoff above 13~GeV. 

In the best-fit ECPL model, the total photon flux above 100~MeV from the SMC is $\Phi_{\gamma}^{>100~{\rm MeV}} = (4.8\pm0.1\pm0.1)\times10^{-8}$~ph~cm$^{-2}$~s$^{-1}$, corresponding to a total energy flux above 100~MeV of $F_{\gamma} = (3.0^{+0.2}_{-0.3}\pm0.1)\times10^{-11}$~erg~cm$^{-2}$~s$^{-1}$. For comparison, this flux estimate is $\sim$30\% greater than that estimated by \cite{abdo10b}, who found (in their ECPL fit) $\Phi_{\gamma}^{>100~{\rm MeV}} = (3.7\pm0.7)\times10^{-8}$~ph~cm$^{-2}$~s$^{-1}$ and is consistent with the flux estimate of (4.7$\pm$0.7)$\times10^{-8}$~ph cm$^{-2}$~s$^{-1}$ from \cite{caputo16}. The large difference between our value and that of \cite{abdo10b} can be attributed to the higher-energy cutoff of our model using $\sim6\times$ more data. Assuming a distance of $D = 61$~kpc to the SMC, the energy flux derived above corresponds to a luminosity of $L_{\gamma} = (1.3\pm0.1\pm0.1)\times10^{37}$ erg~s$^{-1}$. This gamma-ray luminosity remains the lowest to date among star-forming galaxies that have been detected by {\it Fermi} (e.g., \citealt{ackermann12}). 

\begin{deluxetable*}{lccccccc}
\tablecolumns{8}
\tablewidth{0pt} \tablecaption{Spectral Fits Results \label{table:spectra}} 
\tablehead{\colhead{Model} & \colhead{Index~1} & \colhead{Index~2} & \colhead{Break} & \colhead{Cutoff} & \colhead{$\Phi_{\gamma}^{>100~{\rm MeV}}$} & \colhead{log$\mathcal{L}$} & \colhead{TS}} 
\startdata
Power Law & $-$2.27$\pm$0.03$\pm$0.03 & -- & -- & -- & 5.4$\pm$0.2$\pm$0.2 & $-$84029.1 & 1080.4 \\
Broken Power Law & $-$2.10$\pm$0.07$^{+0.06}_{-0.08}$ & $-$2.44$\pm$0.0.7$\pm$0.01 & 1.1$\pm$0.2$\pm$0.1 & -- & 4.7$^{+1.8}_{-1.4}\pm$0.5 & $-$84024.9 & 1088.8 \\
Exponential Cutoff & $-$2.11$\pm$0.06$\pm$0.06 & -- & -- & 13.1$\pm$5.1$\pm$1.6 & 4.8$\pm$0.1$\pm$0.1 & $-$84023.4 & 1091.8 
\enddata
\tablecomments{Columns from left to right: spectral model for the SMC, the spectral index~1, the spectral index~2 (for the broken power law model), the spectral break in GeV (for the broken power law), the cutoff energy in GeV (for the exponential cutoff model), the photon flux in the 100~MeV to 500~GeV band in $10^{-8}$~ph~cm$^{-2}$~s$^{-1}$, fit likelihood, and the TS value for the fit.}
\vspace{-7mm}
\end{deluxetable*}

\subsection{NGC~362} \label{sec:ngc362}

In our likelihood analysis, we found that the addition of a point source at the location of the Galactic globular cluster NGC~362 ($\alpha=15.65^{\circ}$ and $\delta = -70.94^{\circ}$) improved the fit. This point source is directly north of the SMC in the $2-300$ GeV images in Figure~\ref{fig:fermi_image} (where it is labeled for reference) and in the TS map in Figure~\ref{fig:tsmap}. 

For the spectral model of NGC~362, we assume an exponentially cutoff power-law, and we obtain a value of ${\rm TS} = 32$. In this case, the best-fit photon index and cutoff energy were $\Gamma$=1.0$\pm$0.8$^{+1.2}_{-0.8}$ and $E_{ \rm c}$ = 1.6$\pm$1.0$^{+16.6}_{-0.6}$~GeV, respectively.  These values are consistent with those from other {\it Fermi}-detected globular clusters \citep{abdo_gc}. 

We estimate that the photon flux ($>$100~MeV) from NGC~362 is \hbox{$\Phi_{\gamma}^{>100~{\rm MeV}} = (9.4^{+8.7+19.6}_{-3.8-1.9})\times10^{-10}$~ph~cm$^{-2}$~s$^{-1}$}, corresponding to an energy flux ($>$100~MeV) of $F_{\gamma} = (1.0^{+1.0+12.9}_{-0.7-0.4})\times10^{-12}$~erg cm$^{-2}$~s$^{-1}$. The latter value is slightly above the 2$\sigma$ energy flux upper-limit found by \cite{hooper16} of $F_{\gamma} < 8.91\times10^{-13}$~erg~cm$^{-2}$~s$^{-1}$ in the 0.1--100~GeV band using 85 months of {\it Fermi}-LAT data\footnote{We note that \cite{hooper16} adopted the spatial and spectral models of the SMC given in the 3FGL. As we have improved upon those SMC models in our analysis, the results presented here for NGC~362 are likely more reliable than those of \cite{hooper16}.}.  Assuming a distance of 8.5~kpc to NGC~362 \citep{paust10}, the derived energy flux here corresponds to a luminosity of $L_{\gamma}=(8.6^{+8.6+111}_{-6.0-3.4})\times10^{33}$~erg~s$^{-1}$. This luminosity is slightly below (but consistent within the uncertainties) the values of the 15 Galactic globular clusters in the 3FGL \citep{3FGL} which e.g., span a range in luminosity of $L_{\gamma} \sim$ (1--40)$\times10^{34}$~erg s$^{-1}$. 

From $L_{\gamma}$, it is possible to estimate the number of millisecond pulsars (MSPs) in a globular cluster $N_{\rm MSP}$ using the relation

\begin{equation}
N_{\rm MSP} = \frac{L_{\gamma}}{\langle \dot{E} \rangle \langle \eta_{\gamma} \rangle}
\end{equation}

\noindent
where $\langle \dot{E} \rangle$ is the average spin-down power of MSPs and $\langle \eta_{\gamma} \rangle$ is the average spin-down to gamma-ray luminosity conversion efficiency. Following the assumptions of \cite{abdo_gc} that $\langle \dot{E} \rangle = (1.8\pm0.7)\times10^{34}$~erg s$^{-1}$ and  $\langle \eta_{\gamma} \rangle = 0.08$ (see their Section~3.2), NGC~362 has \hbox{$N_{\rm MSP} = 6^{+6+77}_{-4-2}$}, fewer than the globular clusters reported in \cite{abdo_gc}, although the error bars are quite large.  

Previous work has noted a linear correlation between $N_{\rm MSP}$ (or $L_{\gamma}$) and the stellar encounter rate $\Gamma_{\rm e}$ in globular clusters \citep{abdo_gc}: 

\begin{equation}
N_{\rm MSP} = 1.5\times10^{-5} (0.5\pm0.2)\Gamma_{\rm e}+(18\pm9).
\label{eq:encounter}
\end{equation}

\noindent
To explore whether NGC~362 is consistent with this relation, we compute $\Gamma_{\rm e}$ using $\Gamma_{\rm e} = \rho_{0}^{1.5} r_{\rm c}^2$, where $\rho_{0}$ is the central cluster density (in units of $L_{\sun}$~pc$^{-3}$) and $r_{\rm c}$ is the cluster core radius (in pc). We adopt $\rho_{0} = 5.6\times10^{4}$~$L_{\sun}$~pc$^{-3}$ and $r_{\rm c} = 0.45$~pc (from the December 2010 revision of the \citealt{harris96} catalog\footnote{http://physwww.mcmaster.ca/$\sim$harris/mwgc.ref}), and we find $\Gamma_{\rm e} = 2.6\times10^{6}~L_{\sun}^{1.5}~{\rm pc}^{-2.5}$. Using Equation~\ref{eq:encounter}, we derive $N_{\rm MSP} = 37\pm12$. This value is greater than our estimate of $N_{\rm MSP}$ above, although the two numbers are consistent given the large errors. 

\section{Discussion} \label{sec:discussion}

\subsection{Gamma-ray Emissivity of the SMC} \label{sec:emissivity}

Using deep {\it Fermi} data, we have demonstrated that the $2-300$~GeV gamma-ray emission from the SMC has substantial substructure that correlates with the star-forming Bar and Wing. Additionally, its integrated gamma-ray spectrum has a power-law slope of $\Gamma \approx -2.1$ below $\sim$13~GeV, while it is not detected at energies $\gtrsim$13~GeV. Consequently, the best-fit, single component model of the gamma-ray spectrum is a power-law with an exponential cutoff at $\sim$13~GeV. 

For comparison, other star-forming galaxies that have been detected with {\it Fermi} have gamma-ray luminosities much greater than that of the SMC, ranging from $L_{\gamma} = (4.7\pm0.5)\times10^{37}$~erg~s$^{-1}$ (the Large Magellanic Cloud) to \hbox{$L_{\gamma} = (1.5\pm0.6)\times10^{41}$~erg~s$^{-1}$} (NGC~1068; see e.g., \citealt{abdo10d,ackermann12,sudoh18}). None shows an energy cutoff in their gamma-ray spectra, and all except the MW (as discussed below) are best-fit with a single power-law of index $\Gamma \approx -2.2$ \citep{ackermann12}.

In their original SMC {\it Fermi}/LAT analysis, \cite{abdo10b} noted that it is possible that a large fraction of the SMC's diffuse gamma-ray emission arose from unresolved sources, particularly gamma-ray pulsars. Assuming all young pulsars are gamma-ray emitters for 0.1~Myr, \cite{abdo10b} estimated there could be 51$\pm$36 gamma-ray pulsars in the SMC, based on the prediction by \cite{crawford01} that the SMC has 5100$\pm$3600 active radio pulsars with mean lifetimes of 10~Myr. Assuming the pulsars each have values consistent with the median luminosity of young, {\it Fermi}-detected pulsars of $\sim8\times10^{34}$~erg~s$^{-1}$ \citep{abdo13}, the total luminosity from 51$\pm$36 gamma-ray pulsars would be (4.1$\pm2.9)\times10^{36}$~erg~s$^{-1}$, $\sim$30\% of the total luminosity observed from the SMC. 

\begin{figure}
\begin{center}
\includegraphics[width=\columnwidth]{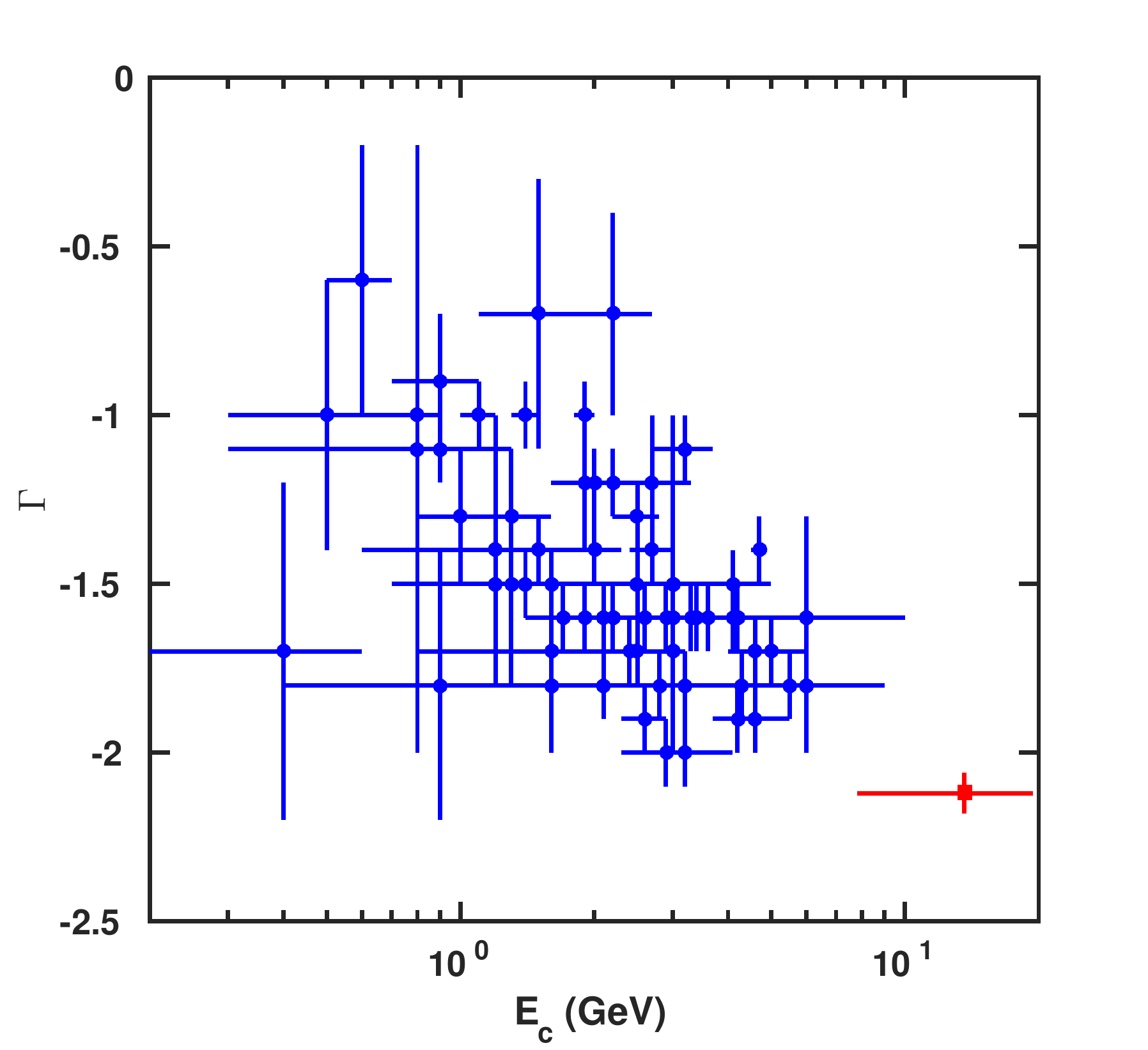}
\end{center}
\vspace{-3mm}
\caption{Best-fit spectral parameters of Milky Way $\gamma$-ray pulsars (blue circles; from Table~9 of \citealt{abdo13}) compared to the integrated spectral parameters of the SMC (red square).}
\label{fig:pulsars}
\end{figure}

High-energy cutoffs are common in the spectra of gamma-ray pulsars, but the best-fit values of $\Gamma$ and $E_{\rm c}$ in the ECPL model of the SMC are not consistent with those typically found from that population. For example, \cite{abdo13} reported spectral characteristics of 70~MW gamma-ray pulsars (see Figure~\ref{fig:pulsars}), which had a median $\Gamma=-1.6$ and $E_{\rm c}=2.5$~GeV with standard deviations of $\Delta \Gamma = 0.3$ and $\Delta E_{\rm c} = 1.4$~GeV, respectively. Thus, the high-energy cutoff of $E_{\rm c} = 13.1\pm5.1\pm1.6$~GeV in the ECPL model of the SMC spectrum cannot be accounted for by pulsars alone. 

We note that some globular clusters with collections of gamma-ray bright, millisecond pulsars have $E_{\rm cut} \sim 10$~GeV (\citealt{hooper16}; though the uncertainties on these values may be large as they were not calculated in that work). We note that none of the brightest globular clusters with luminosities comparable to the SMC have $E_{\rm cut}$ values consistent with $\sim$13~GeV. Additionally, given that the gamma-ray morphology does not follow the distribution of old stars (as shown in Figure~\ref{fig:oldstars}), it is unlikely that a similar population of unresolved millisecond pulsars is powering the observed gamma-ray emission in the SMC.

To explore the contribution of pulsars further, we tested whether the spectrum could be fit by two components. We employed a ECPL$+$BPL model, where the former component represents the contribution from pulsars and the latter component represents the pion decay associated with CRs. To limit the number of free parameters, we froze the ECPL model to have $\Gamma = -1.6$ and $E_{\rm cut} = 2.5$~GeV, consistent with the spectral properties of MW gamma-ray pulsars \citep{abdo13}. We then performed multiple fits, changing the normalization of the ECPL component to assess which pulsar contribution to the total flux best described the data. 

\begin{figure}[t!]
\begin{center}
\includegraphics[width=\columnwidth]{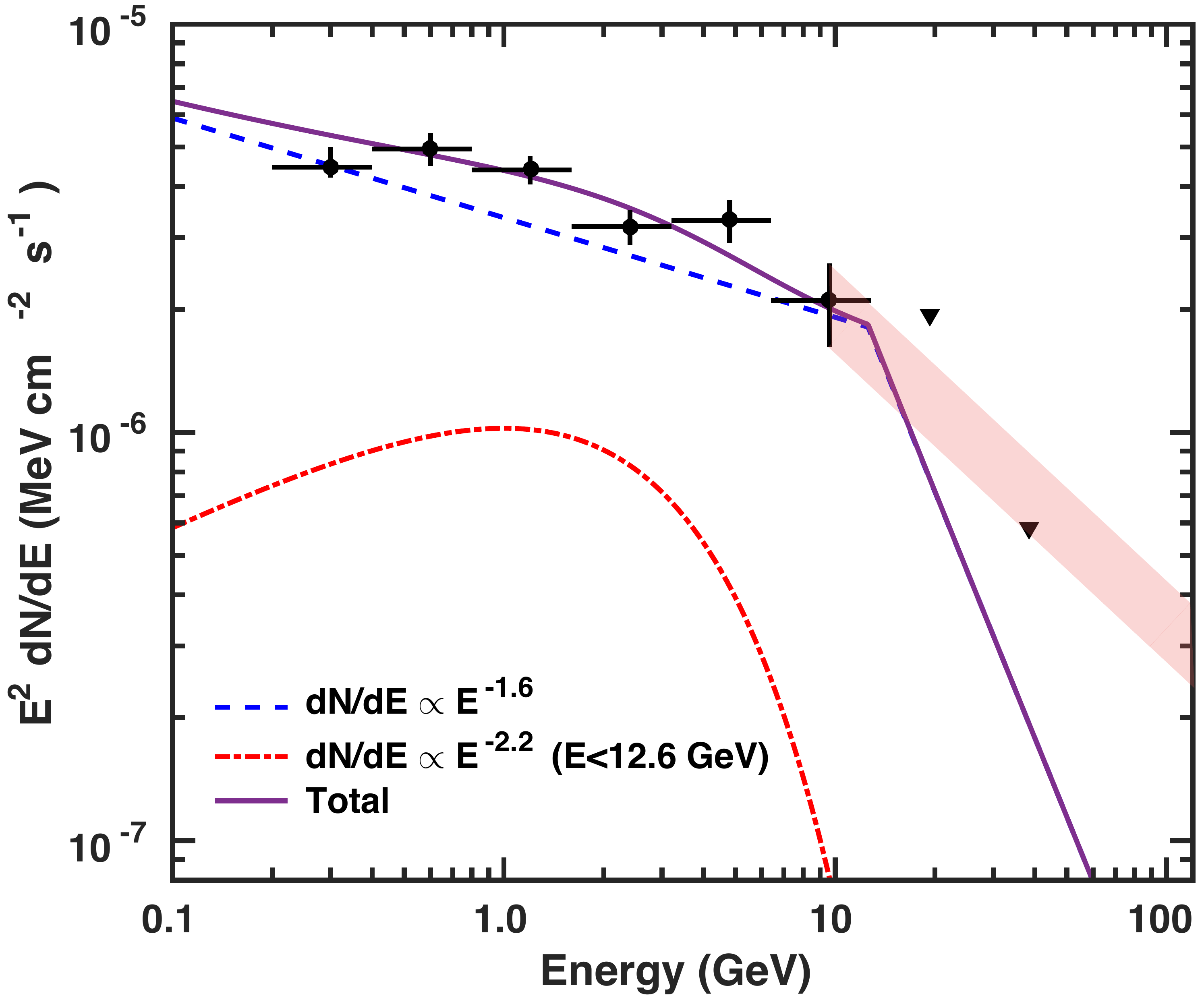}
\end{center}
\vspace{-5mm}
\caption{Gamma-ray spectrum of the SMC with the best-fit two component (ECPL$+$BPL) model overplotted. The blue dashed line represents the ECPL component (with $\Gamma = -1.6$ and $E_{\rm cut} = 2.5$~GeV) from pulsars, the red dash-dotted line is the BPL component from cosmic rays, and the purple solid line is the total of the two components. In this case, the pulsars contribute 14$^{+4}_{-2}$\% of the total photon flux  $\Phi_{\gamma}^{>100~{\rm MeV}}$ above 100~MeV. The red shaded region corresponds to a spectral index of $\Gamma_{2} = -2.75$, as would be expected if high-energy CRs are diffusively escaping from the SMC.}
\label{fig:spectra_twocomp}
\end{figure}

From this set of fits, the best model is plotted in Figure~\ref{fig:spectra_twocomp}. The best-fit BPL has a break at $12.6\pm0.4^{+0.1}_{-0.2}$~GeV and spectral indices of $\Gamma_{1} = -2.24\pm0.2^{+0.05}_{-0.08}$ below and $\Gamma_{2} < -3.9$ above the break energy. In this model, the total photon flux is $\Phi_{\gamma}^{>100~{\rm MeV}} = (5.5^{+0.9}_{-1.2})\times10^{-8}$~ph~cm$^{-2}$~s$^{-1}$, and the BPL (ECPL) components, which represent the CRs (pulsars), contribute 86$^{+2}_{-4}$\% (14$^{+4}_{-2}$\%) to the total. We note that the break energy is the same (within the uncertainties) as the cutoff energy in the single-component ECPL fit from above. Although the spectral index above the break energy is under-constrained due to the upper limits above 12.8 GeV, the fit suggests that the spectrum may steepen at energies $\gtrsim$13~GeV. Statistically, the two-component fit is only marginally better (with $\Delta {\rm TS} = 0.8$) than the best-fit single ECPL model reported in Table~\ref{table:spectra}. However, it demonstrates the plausibility that pulsars may contribute non-negligibly ($\lesssim$20\%) to the total gamma-ray emission from the SMC, though the gamma-rays are likely predominantly produced by CRs.

Consequently, we interpret the morphology and spectrum of the $2-300$~GeV photons as evidence of the CR population in the SMC. In this case, the non-detection of gamma-rays above $\gtrsim$13~GeV may indicate that the spectrum steepens because of diffusive escape of CR protons from the SMC halo, as discussed in Section~\ref{sec:escape} below.

Assuming that the gamma-ray emission does arise from CRs interacting with interstellar gas, the integrated $>$100~MeV gamma-ray emissivity per hydrogen atom $q_{\gamma}^{>100~{\rm MeV}}$ can be calculated using

\vspace{-1mm}
\begin{equation}
q_{\gamma}^{>100~{\rm MeV}} = \Phi_{\gamma}^{>100~{\rm MeV}} \frac{m_{\rm p}}{M_{\rm gas}} D^2
\label{eq:q}
\end{equation}

\noindent
where $\Phi_{\gamma}^{>100~{\rm MeV}}$ is the integrated photon flux above 100~MeV, $M_{\rm gas}$ is the total gas mass of the galaxy, and $D$ is the distance to the SMC, $D = 61$~kpc. The SMC gas mass is dominated by atomic hydrogen, with a mass $M_{\rm HI} = 4.2\times10^{8}~M_{\sun}$ \citep{stan99}, whereas the molecular hydrogen (H$_{2}$) gas mass is estimated to be lower, $M_{\rm H_{2}} \approx (1-3)\times10^{7}~M_{\sun}$ \citep{leroy07,bolatto11,jameson16}. Thus, the total gas mass is $M_{\rm gas} = M_{\rm HI} + M_{\rm H_{2}} \approx (4.3-4.5)\times10^{8}~M_{\sun}$. Using Equation~\ref{eq:q}, we find $q_{\gamma}^{>100~{\rm MeV}} = (3.1\pm0.1)\times10^{-27}$~ph~s$^{-1}$~sr$^{-1}$~H$^{-1}$, assuming 86\% of the photon flux $\Phi_{\gamma}^{>100~{\rm MeV}} = (5.5^{+0.9}_{-1.2})\times10^{-8}$~ph~cm$^{-2}$~s$^{-1}$ from the two-component spectral model arises from CRs. 

In our emissivity calculation, we have assumed that all of the gamma-rays are from $\pi^{0}$ decay (rather than leptonic processes) and that unresolved pulsars contribute 14\% to the total photon flux (from the analysis above). Thus, this $q_{\gamma}^{>100~{\rm MeV}}$ should be viewed as an upper limit on the emissivity. We note that our derived $q_{\gamma}^{>100~{\rm MeV}}$ for the SMC is 25\% greater than that reported by \cite{abdo10b} since our $\Phi_{\gamma}$ is larger.

By comparison, the average gamma-ray emissivity of the Milky Way ISM is $\sim$5 times greater, with $q_{\gamma}^{>100~{\rm MeV}} = (1.63\pm0.05)\times10^{-26}$~ph~s$^{-1}$~sr$^{-1}$~H$^{-1}$ \citep{abdo09}. The low emissivity of the SMC suggests that the average density of CR nuclei in the SMC is $\sim$5$\times$ less than in the MW. Assuming diffusive shock acceleration operates similarly between galaxies, then this result would arise from either a lower CR injection rate per unit star-forming volume\footnote{The FIR/radio correlation \citep{helou85,condon92,yun01} suggests that the efficiency of producing CR electrons per unit star formation is constant from galaxy to galaxy, assuming that all GHz radio emission from star-forming galaxies results from synchrotron cooling of CR electrons and that the FIR emission is due to reprocessed starlight onto dust \citep{socrates08}.} $\dot{E}_{\rm CR}/V$ or from a smaller confinement length $l_{\rm conf}$ in the SMC. 

\subsection{Escape of Cosmic Rays from the SMC} \label{sec:escape}

Galaxies are ``calorimeters" of CR protons when all accelerated CR protons experience pion losses, as in e.g., starburst galaxies \citep{thompson07,socrates08,lacki11,ackermann12}. To assess how close the SMC is to this calorimetric limit, we estimate the ratio of the observed gamma-ray luminosity $L_{\gamma}$ to the maximum gamma-ray luminosity $L_{\gamma}^{\rm max}$ possible given the CR injection rate $\dot{E}_{\rm CR}$. Here we define this calorimetry fraction as $f_{\rm cal} \equiv L_{\gamma} / L_{\gamma}^{\rm max}$. 

The CR injection rate is 

\begin{equation}
\dot{E}_{\rm CR} = \eta E_{\rm SN} \Gamma_{\rm SN},
\label{eq:ECR}
\end{equation}

\noindent
where $\eta$ is the fraction of the supernova (SN) kinetic energy that goes into primary CR protons, $E_{\rm SN}$ is the SN kinetic energy, and $\Gamma_{\rm SN}$ is the rate of SNe in the SMC. We assume $\eta = 0.1$, $E_{\rm SN} = 10^{51}$~erg, and $\Gamma_{\rm SN} = 0.0015~{\rm yr}^{-1}$. The latter quantity is derived by multiplying the MW SN rate of 0.02~yr$^{-1}$ by the ratio of the star formation rates (SFRs) in the SMC ($\sim$0.1~$M_{\odot}$~yr$^{-1}$: \citealt{harris04}) to that of the MW ($\sim$1.3~$M_{\odot}$~yr$^{-1}$: \citealt{murray10,rob10}). This $\Gamma_{\rm SN}$ is consistent with the known supernova remnant (SNR) population in the SMC \citep{badenes10,auchettl18} if their visibility time is $\sim$15,000 years (near the expected visibility time of $\sim$20,000 years from semi-analytic modeling: \citealt{sumit17}). Using the above values, we find $\dot{E}_{\rm CR} = 4.75\times10^{39}$~erg~s$^{-1}$ for the SMC. 

The maximum gamma-ray luminosity that can be produced by this CR injection rate is $L_{\gamma}^{\rm max} = f_{\gamma} \dot{E}_{\rm CR}$, where $f_{\gamma} = 1/3$ is the the fraction of pions that decay to gamma-rays. Therefore, we find \hbox{$L_{\gamma}^{\rm max} = 1.6\times10^{39}$~erg~s$^{-1}$}, and thus $f_{\rm cal} = 0.007$, given the observed gamma-ray luminosity from CRs of \hbox{$L_{\gamma} = (1.1\pm0.1)\times10^{37}$~erg~s$^{-1}$} (using 86\% of the total luminosity from Section~\ref{sec:SMC}). By comparison, the MW has $f_{\rm cal} = (7\times10^{38}~{\rm erg~s}^{-1})/(2.1\times10^{40}~{\rm erg~s}^{-1}) =$ 0.033 \citep{strong10,ackermann12}. 

In the MW, the small $f_{\rm cal}$ is attributed to CRs escaping diffusively from the galaxy's halo, since the CR diffusion time $t_{\rm diff} \approx 45 (\frac{E_{\rm CR}}{1~{\rm GeV}})^{-1/2}$~Myr is less than the pion loss timescale, $t_{\pi} \approx 100 (\frac{n_{\rm eff}}{0.5~{\rm cm}^{-3}})^{-1}$~Myr \citep{mannheim94}. In these relations, $E_{\rm CR}$ is the CR energy, and $n_{\rm eff}$ is the effective density encountered by the CRs (in the MW, $n_{\rm eff}\approx$ 0.2--0.5 cm$^{-3}$: \citealt{cornell98,sch02}). For comparison, the advective escape timescale is $t_{\rm adv} = h / v_{\rm wind} \approx 10 (\frac{h}{1~{\rm kpc}}) (\frac{v_{\rm wind}}{100~{\rm km~s}^{-1}})^{-1}$~Myr, where $h$ is the galaxy's scale height and $v_{\rm wind}$ is the galactic wind velocity.

As per the calculation above, we find $f_{\rm cal}$ of the SMC is $\sim5\times$ smaller than that of the MW. We caution that there are large uncertainties in $\dot{E}_{\rm CR}$ (i.e., the SFRs), and we have assumed that the SMC's $L_{\gamma}$ is produced exclusively by pion decay associated with CR protons. Thus, our derived $f_{\rm cal}$ for the SMC is likely an upper limit, given that CR electrons may contribute non-negligibly to the spectrum. If the SMC's $f_{\rm cal}$ is indeed lower than that of the MW, then it could be either due to more escape of CRs (through diffusion or advection) or from fewer pionic losses than in the MW. The former explanation could result from a smaller confinement length or larger diffusion coefficient $D_{0}$, since $t_{\rm diff} \sim l^{2}_{\rm conf}/D_{0}$, where $D_{0}$ is the diffusion coefficient. Alternatively, the SMC could have fewer pionic losses than the MW if $n_{\rm eff}$ is lower in the SMC than in the MW. However, we expect that the pion loss timescale of the SMC is comparable to the MW, given that the SMC has \hbox{$n_{\rm eff} \sim 0.2$~cm$^{-3}$}, assuming a median hydrogen column density of $N_{\rm H} = 2\times10^{21}$~cm$^{-2}$  \citep{stan99} and a depth of $\sim$4~kpc \citep{muraveva18}.

In the MW where CR proton lifetimes are set by diffusive escape, the GeV to PeV proton spectra go as $E^{-2.75}$ \citep{simpson83,sanuki00,adriani11}. By contrast, if CRs experience pionic losses or escape via advection, spectra can be harder and go as $E^{-2}$--$E^{-2.4}$ (as in e.g., M82 and NGC~253: \citealt{lacki11,ackermann12}). Thus, the best-fit spectral models for the SMC plotted in Figures~\ref{fig:spectra} and \ref{fig:spectra_twocomp} are consistent with CR proton lifetimes limited by pionic losses or advection. However, given the sub-calorimetric luminosity of the SMC from above, it is apparent that the CRs are not being efficiently converted to gamma-rays. 

Consequently, the luminosity and spectrum of the SMC is most consistent with the scenario where advection sets the spectrum $\lesssim$13~GeV and diffusive losses produce a steeper spectrum $\gtrsim$13~GeV. In this case, the cutoff energy in the best-fit, single component ECPL model could be suggestive of the transition in the spectrum from advection- to diffusion-dominated. The energy break in the best-fit ECPL$+$BPL model of Section~\ref{sec:emissivity} may be interpreted similarly. 

In the latter model, $\Gamma_{2}$ is much steeper than the $E^{-2.75}$ spectrum observed in the MW. However, $\Gamma_{2}$ is not well constrained given the lack of a statistically significant detection in the two highest energy bins. In Figure~\ref{fig:spectra_twocomp}, the red shaded region represents a $\Gamma_{2} = -2.75$ spectrum above the $6.4-12.8$~GeV data point. This $\Gamma_{2}$ is consistent with our upper limits $>$13~GeV if the energy flux in the $6.4-12.8$~GeV band is toward the lower bound of the error bar. As {\it Fermi} continues to collect data from the SMC, increased count statistics above 13~GeV will reveal whether our interpretation of the cutoff as spectral steepening is correct.

To date, no detection of a wind from the SMC has been reported in the literature that is consistent with advective losses. H{\sc i} observations do show multiple expanding, supergiant shells with velocities of $\sim$30~km~s$^{-1}$ \citep{stan99}.

We can make a rough estimate of the confinement length $l_{\rm conf}$ in the SMC assuming that the CR protons of energy $E_{\rm CR} = 130$~GeV (corresponding to the 13~GeV photons in the spectrum) are escaping diffusively. For this calculation, we adopt two diffusion coefficients spanning a range found observationally: $D_{0} = 10^{27}$~cm$^{2}$~s$^{-1}$ (as obtained near the star-forming region 30 Doradus: \citealt{murphy12}) and $D_{0} = 5\times10^{28}$~cm$^{2}$~s$^{-1}$ (which is found in the MW: \citealt{trotta11}). Given $t_{\rm diff} \approx 45 (\frac{E_{\rm CR}}{1~{\rm GeV}})^{-1/2}$~Myr, $t_{\rm diff} = 4$~Myr for $E_{\rm CR} = 130$~GeV. Solving for $l_{\rm conf}$, we find $l_{\rm conf} \approx$ 110~pc or 800~pc for the two diffusion coefficients, respectively. Thus, even for large diffusion coefficients, $l_{\rm conf}$ is less than the size of the star-forming Bar ($\sim$1~kpc across) and the depth of the SMC ($\sim$4~kpc) for the $\sim$130~GeV CR protons producing the 13~GeV photons.

We note that the statistics of the current {\it Fermi} data do not allow us to explore how the spectrum changes as a function of position across the SMC. In the MW, there is evidence of radial gradients in the efficiency of CR transport (e.g., \citealt{yang16}). If true for the SMC, then the spectra may be harder or softer locally, depending on e.g., the concentration of CR particle accelerators or changes in the diffusion coefficient. In particular, an alternative interpretation of the spectral cutoff is that there may be a steep spatial gradient in the diffusion constant throughout the SMC. In this scenario, low-energy CRs are trapped in low-diffusion regions near their acceleration sites, whereas higher-energy CRs enter areas of greater diffusion constants and can easily escape the galaxy. Models of CR self-confinement near SNRs obtain this phenomenology, with sharp cutoffs in the CR confinement time ~\citep{ptuskin08,dangelo18}. This scenario has been invoked to explain the hard gamma-ray spectra of MW SNRs~\citep{dangelo18}, and it would also affect the integrated spectrum observed from a galaxy~\citep{evoli18}. In the future, deeper {\it Fermi} data will enable comparison of the spectra across multiple locations in the SMC to explore this interpretation.

\section{Conclusions} \label{sec:conclusions}

We have analyzed 105~months of {\it Fermi} data toward the Small Magellanic Cloud, and we have presented $2-300$~GeV images that have substantial substructure correlated with the star-forming Bar and Wing of the SMC. The SMC is not detected above $\sim$13~GeV, and we set strict upper-limits on the flux at these energies. A simple power-law model is inadequate at describing the SMC's GeV spectrum, and a power-law with an exponential cutoff at $\sim$13 GeV is statistically significantly better. We perform two-component fits to assess the relative contribution of pulsars and CRs to the emission, and we find that pulsars contribute 14$^{+4}_{-2}$\% to the total flux above 100~MeV. In this case, the CR component has a hard spectral index of $\sim-2.2$ below $\sim$12.6~GeV and steepens substantially at higher energies.

We show that the gamma-ray emissivity of the SMC is $\sim$5$\times$ less than that of the MW, and the SMC's gamma-ray luminosity is only $\sim$0.7\% of the maximum possible luminosity given the CR injection rate in the SMC (the ``calorimetric limit"). In conjunction with the spectral results, we attribute these characteristics to the advective and diffusive escape of CRs from the SMC. In this scenario, the gamma-ray spectrum is harder below $\sim$13~GeV because CR protons producing those photons have lifetimes set by advection, whereas above that limit, the CR protons are lost via energy-dependent diffusive escape. In the future, increased photon statistics above $\sim$13~GeV with deeper {\it Fermi} data are necessary to determine whether the exponential cutoff reported here is actually a steep spectrum indicative of diffusive losses.

\acknowledgements

We thank Daniel Castro and Dennis Zaritsky for useful discussions that improved the quality of this paper. L.A.L. acknowledges support from the Sophie and Tycho Brahe Visiting Professorship at the Niels Bohr Institute, and E.R.R. acknowledges support from the DNRF for the Niels Bohr Professorship.

\nocite{*}
\bibliographystyle{aasjournal}
\bibliography{cr_smc}

\end{document}